\documentclass[prx,10pt,floatfix,twocolumn,notitlepage]{revtex4-1}

\usepackage{graphicx}
\usepackage{dcolumn}
\usepackage{amsmath}
\usepackage{amsthm}
\usepackage{amssymb}
\usepackage{physics}
\usepackage{bm}
\usepackage{color}
\usepackage{times}
\usepackage[normalem]{ulem}
\usepackage[justification=raggedright]{caption}

\usepackage{subcaption}
\usepackage{tabularx}
\usepackage{appendix}

\usepackage{amstext}
\usepackage{array}

\usepackage{diagbox}

\usepackage{bbm}

\definecolor{wine-stain}{rgb}{0.5,0,0}
\definecolor{bblue}{rgb}{0,0.0,0.5}

\usepackage[colorlinks=true
,urlcolor=bblue
,anchorcolor=blue
,citecolor=wine-stain
,filecolor=blue
,linkcolor=blue
,menucolor=blue
,pagecolor=blue
,linktocpage=true
,pdfproducer=medialab
,linktoc=all 
]{hyperref}

\usepackage{multirow}

\usepackage{booktabs}  

\allowdisplaybreaks

\usepackage{setspace}

\usepackage{custom_commands}

\newcommand{\subtxt}[2]{{#1}_\text{\scriptsize{#2} } }
\ncmd{\el}{\subtxt{\mf{l}}{BZ} }

\begin{document}

\title{
Coulomb interaction driven instabilities of sliding Luttinger liquids
}

\author{Shouvik Sur and Kun Yang}
\affiliation{National High Magnetic Field Laboratory and Department of Physics,\\
Florida State University, Tallahassee, Florida 32306, USA
}

\date{\today}

\begin{abstract}
We study systems made of periodic arrays  of one dimensional quantum wires coupled by Coulomb interaction.
Using bosonization an interacting metallic fixed point is obtained, which is shown to be a higher dimensional analogue of the Tomonaga-Luttinger liquid, or a sliding Luttinger liquid.
This non-Fermi liquid metallic state, however, is unstable in the presence of weak interwire backscatterings, which favor charge density wave states and suppress pairing.
Depending on the effective strength of the Coulomb repulsion and the size of interwire spacing various charge density wave states are stabilized, including Wigner crystal states. 
Our method allows for the determination of the specific ordering patterns, and corresponding energy and temperature scales.
\end{abstract}

\maketitle



\section{Introduction} \label{sec:intro}
Tomonaga-Luttinger liquid (or simply Luttinger liquid, abbreviated as LL) is the generic metallic state realized in interacting one dimensional systems  \cite{Tomonaga50, Luttinger63, Mattis65, Haldane81}.
Phase space constraints and nesting enable  weak short-range interactions to destroy quasiparticle coherence while preserving  metallicity, making LL the earliest example of a metal that is not described by Fermi liquid theory.
Thus, LLs have long motivated theoretical modeling of  non-Fermi liquid states in higher dimensions \cite{Fabrizio93-a, Fabrizio93-b, Finkelstein93, Anderson94}.
Indeed within a coupled-wire construction it was shown that a higher-dimensional analogue of LL, the sliding Luttinger liquid (SLL), can be stabilized above one dimension in the presence of short-range repulsive interactions and/or vanishing interwire hoppings \cite{Emery00, Vishwanath01, Mukhopadhyay01-a, Sondhi01}. Despite the coupling between wires, the SLL possesses an emergent `sliding' symmetry corresponding to {\em independent} translation invariance on each wire,  whose nature  will be made precise in section \ref{sec:model}.
The SLL is an anisotropic metal, which behaves like a LL along the wires, while transport is suppressed in the transverse direction(s).

In parallel to these, coupled LLs have also been used as a paradigm to study competing orders above one dimension \cite{Kivelson98, Fradkin99}.
In one dimension, strong quantum fluctuations prevent spontaneous breaking of continuous symmetries  \cite{yang04}.
Instead tendencies toward ordering in different channels manifest themselves in the power-law correlations of various local order parameters, with the power law exponents (or scaling dimensions) indicating the strength of the ordering tendency.
Indeed the {\em unstable} coupled LL fixed point can be used as the starting point of systematic analyses of the physics of ordering in coupled LLs, based on renormalization group (RG) arguments.
In recent examples, the paradigm of  wires coupled by short \cite{Bulmash17} and long \cite{Zhang16} ranged interactions was applied to understand the physics of magnetic field driven catalysis in metals with low carrier density  \cite{Celli65,Fukuyama78,Miransky15}.
Quenching of the kinetic energy  on the plane perpendicular to the applied field makes the metal susceptible to density wave ordering.
Owing to a small or vanishing Fermi energy,  the quantum limit is reached at moderate magnetic field strength, and the lowest Landau level dominates the low energy physics.
The degeneracy of the lowest Landau level is utilized to map the problem to that of coupled wires, where the number of wires is controlled by the degeneracy.

In this work we revisit the problem of quantum wires coupled by Coulomb interaction, in the regime where single-electron interwire hoppings are suppressed.
The purpose of our work is the  following. In reality the interwire couplings that lead to SLL physics and those that lead to charge density wave (CDW) ordering  come from the {\em same} Coulomb interaction. They should, therefore, be treated on equal-footing. We will demonstrate that such a treatment leads to specific predictions of the leading CDW instability and resultant ordering pattern, as well as corresponding energy scales.

The paper is organized as follows. In section \ref{sec:model} we introduce the model, and derive the bosonized action for a system of infinite number of quantum wires in $d$ dimensions coupled by  Coulomb interaction in the forward scattering channel.
In section \ref{sec:SLL} we show that the action describes a SLL fixed point in $d$ dimensions, and deduce the  exponents that characterize the fixed point.
In section \ref{sec:instabilities} we analyze the stability of the SLL fixed point against various  symmetry breaking  perturbations.
Within a tree-level RG analysis we show that multiple charge density wave (CDW) states compete for dominance in the absence of a dimensional crossover, and interwire pairing instabilities are suppressed.
We establish the zero (finite) temperature phase diagram as a function of Coulomb interaction strength (temperature) and interwire spacing.
Finally, we close with a discussion of our results in section \ref{sec:conclusion}.

\section{Model} \label{sec:model}
We consider a $d-1$ dimensional lattice of identical wires of spinless fermions in $d$ dimensions with the wires lying  along the $\hat{x}$-axis \cite{note1}.
The wires are labeled by a $d-1$ dimensional vector, $\mbf{n}$, such that $\mbf{n} \cdot \hat x = 0$.
The fermion field on the $\mbf{n}$-th wire is expressed in terms of the hydrodynamic modes as \cite{Haldane81}
\begin{align}
\psi_{\mbf{n}}(\tau,x) &\approx 
e^{-i\pi \rho_0 x} \psi_{L;\mbf{n}}(\tau,x) 
+ e^{i\pi \rho_0 x} \psi_{R;\mbf{n}}(\tau,x),
\label{eq:psi-full}
\end{align}
where modes carrying momenta of magnitude larger than $\pi \rho_0$ are ignored.
The left and right moving fermions are expressed as,
\begin{align}
& \psi_{L;\mbf{n}}(\tau,x) = e^{i (\vtheta_{\mbf{n}}(\tau, x) - \vphi_{\mbf{n}}(\tau, x))} 
\sqrt{ \rho_0 + \frac{1}{\pi} \dow_x \vphi_{\mbf{n}}(\tau, x)} \nn \\
& \psi_{R;\mbf{n}}(\tau,x) = e^{i (\vtheta_{\mbf{n}}(\tau, x) + \vphi_{\mbf{n}}(\tau, x))} 
\sqrt{ \rho_0 + \frac{1}{\pi} \dow_x \vphi_{\mbf{n}}(\tau, x)}.
\end{align}
Here $\rho_0$ is the mean density,  $\dow_x \vphi_{\mbf{n}}$ is local density modulation of fermions on each wire, and $\vtheta_{\mbf{n}}$ is the phase of the fermion field.
The action for free fermions is given by 
\begin{align}
S_0
&= \half \sum_{\mbf{n}} \int_{(\tau, x)}
\lt[
\frac{2i}{\pi} (\dow_x \vphi_{\mbf{n}}) (\dow_\tau \vtheta_{\mbf{n}}) + \frac{v_F}{\pi} (\dow_x \vphi_{\mbf{n}})^2 \rt. \nn \\
& \qquad \qquad \qquad \qquad \lt.
+ \frac{v_F}{\pi} (\dow_x \vtheta_{\mbf{n}})^2,
\rt]
\label{eq:S0-2}
\end{align}
where $\int_x \equiv \int dx$ and  $v_F = k_F / m$ with $k_F = \pi \rho_0$ being the Fermi wavevector, and $m$ being the mass of the non-interacting fermions.

We introduce an instantaneous  interaction among the fermions,
\begin{align}
S_{I} &= \half \sum_{\mbf{n}, \mbf{m}} \int_{(\tau, x, x')}
V( x - x',\mbf{n} - \mbf{m}) ~ \rho_{\mbf{n}}(\tau, x) ~ \rho_{\mbf{m}}(\tau, x'),
\end{align}
where $\rho_{\mbf{n}}(\tau, x) = \rho_0 + \frac{1}{\pi} \dow_x \vphi_{\mbf{n}}(\tau, x)$  is  the density  on the $\mbf{n}$-th wire.
We assume the wires are uniformly spaced with lattice spacing $\mf{a}$, and define the Fourier components through
\begin{align}
\vphi_{\mbf{n}}(\tau,x) &= \el^{d-1} \int_{-\infty}^{\infty} \frac{dk_0 ~ dk_x}{(2\pi)^2} \nn \\
& \quad \times \int_{BZ} \frac{d^{d-1} K}{(2\pi)^{d-1}} ~ e^{i \tau k_0 + i x k_x + i \mf{a} \mbf{n} \cdot  \mbf{K}} ~ \vphi(k),
\end{align}
where $BZ$ indicates the first Brillouin zone, $\el^{-1}$ is a measure of  linear dimension of $BZ$ such that its volume is $\lt(\frac{2\pi}{ \el} \rt)^{d-1}$, and $ \mbf{K}$ represents points inside $BZ$.
Thus the bosonized action for the interacting theory is given by
\begin{align}
S &= \frac{\el^{d-1}}{2\pi} \lt(\frac{\el}{\mf a} \rt)^{d-1} \int dk
\Bigl[
2i k_0 k_x \vphi(-k) \vtheta(k)  \nn \\
& \qquad
+ V_{\vphi}(\vec{k}) ~k_x^2~ \vphi(-k) \vphi(k) 
+ v_F ~k_x^2~ \vtheta(-k) \vtheta(k)
\Bigr],
\label{eq:S-2}
\end{align}
where $dk = \frac{dk_0 dk_x d^{d-1}K}{(2\pi)^{d+1}}$, $\vec{k} \equiv (k_x,  \mbf{K})$, and $V_{\vphi}(\vec k) = v_F + \pi^{-1} (\el/\mf{a})^{d-1} V(\vec k)$ with $V(\vec k)$ being the Fourier conjugate of $V(x-x',\mbf{n} - \mbf{m})$.
In $d=2$ the lattice geometry dependent ratio $\el/\mf{a}$ equals $1$, while in $d=3$ it equals $1$ and $(\sqrt{3}/2)^{1/2}$ for the square and   triangular lattices, respectively.
From the coordinate space representation of \eq{eq:S-2}, we deduce that the action is invariant under wire dependent shifts $\vphi_{\mbf{n}}(\tau, x) \mapsto \vphi_{\mbf{n}}(\tau, x) + \vphi_{\mbf{n}}^{(0)}$ and $\vtheta_{\mbf{n}}(\tau, x) \mapsto \vtheta_{\mbf{n}}(\tau, x) + \vtheta_{\mbf{n}}^{(0)}$, where $\vphi_{\mbf{n}}^{(0)}$ and $\vtheta_{\mbf{n}}^{(0)}$ are constants.
The former invariance is the  sliding symmetry advertised in section \ref{sec:intro}, and it corresponds to  translation invariance  along each wire.
It allows the fermions on distinct  wires to `slide' with respect to each other.
The latter invariance corresponds to particle number conservation on each wire which prevents single-particle interwire hoppings.
This results in two flat patches of Fermi surface that are nested by the wavevector $2 k_F \hat{x}$.
Generally, such extensive nesting makes the metallic state exceedingly susceptible to weak coupling instabilities.
However, with a suitable choice of short-range interwire interactions, it is possible to stabilize this metallic state  in  $d>1$ \cite{Emery00, Vishwanath01,Mukhopadhyay01-a,Sondhi01}.
Although Eqs. \eqref{eq:S0-2} and \eqref{eq:S-2} are both Gaussian actions, they describe a non-interacting and an interacting fixed point, respectively.
We will elucidate this point further through the computation of scaling exponents in subsequent  sections.

\section{Coulombic Sliding Luttinger Liquid} \label{sec:SLL}
In this section we characterize the fixed point described by the action in \eq{eq:S-2}.
We consider both the interwire and intrawire interactions that arise from unscreened Coulomb interaction among fermions, $V(x-x', \mbf{n} - \mbf{m}) = \frac{\mf{e}^2}{4 \pi \veps \sqrt{(x-x')^2 + \mf{a}^2 |\mbf{n} - \mbf{m}|^2 }}$ with $\mf{e}$ being the  electric charge of the fermions and $\veps$ being the permittivity, such that
\begin{align}
V(\vec{k}) = \frac{2 \mf{e}^2}{4\pi \veps} \sum_{\mbf n} e^{i \mf{a} \mbf n \cdot \mbf K} ~ \mc{K}_0(\mf{a} |\mbf n| k_x ),
\end{align}
where $\mc K_0(x)$ is the modified Bessel function of the second kind, and $\mc K_0(x) \sim - \ln{x}$ ($\mc K_0(x) \sim e^{-x} $) for $x \ll 1$ ($x \gg 1$) \cite{Kopietz97}.
In the limit $\mf{a} |\vec k| \ll 1$, the summand varies slowly as a function of $\mbf n$.
Therefore, in this limit we replace the sum by an integral to obtain
\begin{align}
V_{\vphi}(\vec{k}) \approx v_F
+ \frac{A'_d \mf{e}^2}{4\pi \veps \mf{a}^{d-1}} \frac{(\el/\mf{a})^{d-1}}{(k_x^2 + |\mbf K|^2)^{(d-1)/2}},
\end{align}
where $A'_d > 0$, and $(A'_2, A'_3) = (2, 4)$.

In order to obtain the analogue of Luttinger parameters, we project $V_{\vphi}(\vec{k})$  in the forward scattering channel along the wires by setting $k_x = 0$.
Thus we consider
\begin{align}
V_{\vphi}(\mbf{K}) = v_F +  \frac{A_d ~ \mf{e}^2}{4\pi \veps (\mf{a} |  \mbf{K}|)^{(d-1)}},
\end{align}
where $A_d = A'_d (\el/\mf{a})^{d-1}$,  
as the effective Coulomb interaction for scatterings in the forward scattering channel.
By utilizing the two component basis, $\trans{(\vphi, \vtheta)}$, the propagators for $\vtheta(k)$ and $\vphi(k)$ are easily deduced from \eq{eq:S-2},
\begin{align}
& G_\vtheta(k) =  \frac{\pi (\mf{a}/ \el)^{d-1} V_{\vphi}(\vec{k})}{k_0^2 + v_F V_{\vphi}(\mbf{K}) k_x^2 },
\quad
  G_\vphi(k) = \frac{\pi (\mf{a}/ \el)^{d-1} v_F}{k_0^2 + v_F V_{\vphi}(\mbf{K}) k_x^2 }.
\end{align}
Furthermore, they are correlated as
\begin{align}
\avg{\vphi(-k) \vtheta(k)}
=  \frac{\pi (\mf{a}/ \el)^{d-1} k_0}{ik_x \lt(k_0^2 + v_F V_{\vphi}(\mbf{K}) k_x^2 \rt) }.
\end{align}
The equal-time correlation between the simplest vertex operators,
\begin{align}
& \avg{ e^{i \vtheta_{\mbf{n}}(\tau, x)}
e^{- i  \vtheta_{\mbf{m}}(\tau, x + \Dl x)} }
= \frac{\dl_{\mbf{n},\mbf{m}}}{|\lam_0 \Dl x|^{2\eta_\vtheta(\mf{h}_d)}},
\label{eq:corr-theta} \\
& \avg{ e^{i   \vphi_{\mbf{n}}(\tau, x)}
e^{- i \vphi_{\mbf{m}}(\tau, x + \Dl x)} }
= \frac{\dl_{\mbf{n},\mbf{m}}}{|\lam_0 \Dl x|^{2\eta_\vphi(\mf{h}_d)}},
\label{eq:corr-phi}
\end{align}
where $\lam_0^{-1} > k_F^{-1}$ is the short-distance  cutoff along the wires, and
\begin{align}
\mf{h}_d =  \frac{A_d ~\mf{e}^2}{4 \pi^d  \veps v_F}
\label{eq:h}
\end{align}
is the effective fine structure constant.
We derive the   exponents $\eta_\vphi(\mf h_d)$ and $\eta_\vtheta(\mf h_d)$ for various wire-stacking geometries  in Appendix \ref{app:eta}.
To illustrate general features of these exponents here we quote the results for the square lattice,
\begin{align}
& \eta_\vtheta(\mf{h}_d) = \frac{1}{4}
\int_0^1 d^{d-1}{w} \lt(1 + \frac{\mf{h}_d}{|\mbf{w}|^{d-1}} \rt)^{\half},
\label{eq:eta-theta}\\
& \eta_\vphi(\mf{h}_d) = \frac{1}{4}
\int_0^1 d^{d-1}{w} \lt(1 + \frac{\mf{h}_d}{|\mbf{w}|^{d-1}} \rt)^{-\half},
\label{eq:eta-phi}
\end{align}
where the integrations are over the unit $d-1$ dimensional cube, and $\mbf{w}$ is a vector inside the cube.
Since the integrand in \eq{eq:eta-theta} (\eq{eq:eta-phi}) is larger (smaller) than $1$, $\eta_\vtheta(\mf{h}_d) > \eta_\vtheta(0)$ ($\eta_\vphi(\mf{h}_d) < \eta_\vphi(0)$) for any $h_d > 0$.
Therefore, intrawire phase (density) fluctuation is suppressed (enhanced) compared to that at the non-interacting fixed point as the effective strength of the Coulomb interaction increases.
\begin{figure}[!t]
\centering
\includegraphics[width=0.85\columnwidth]{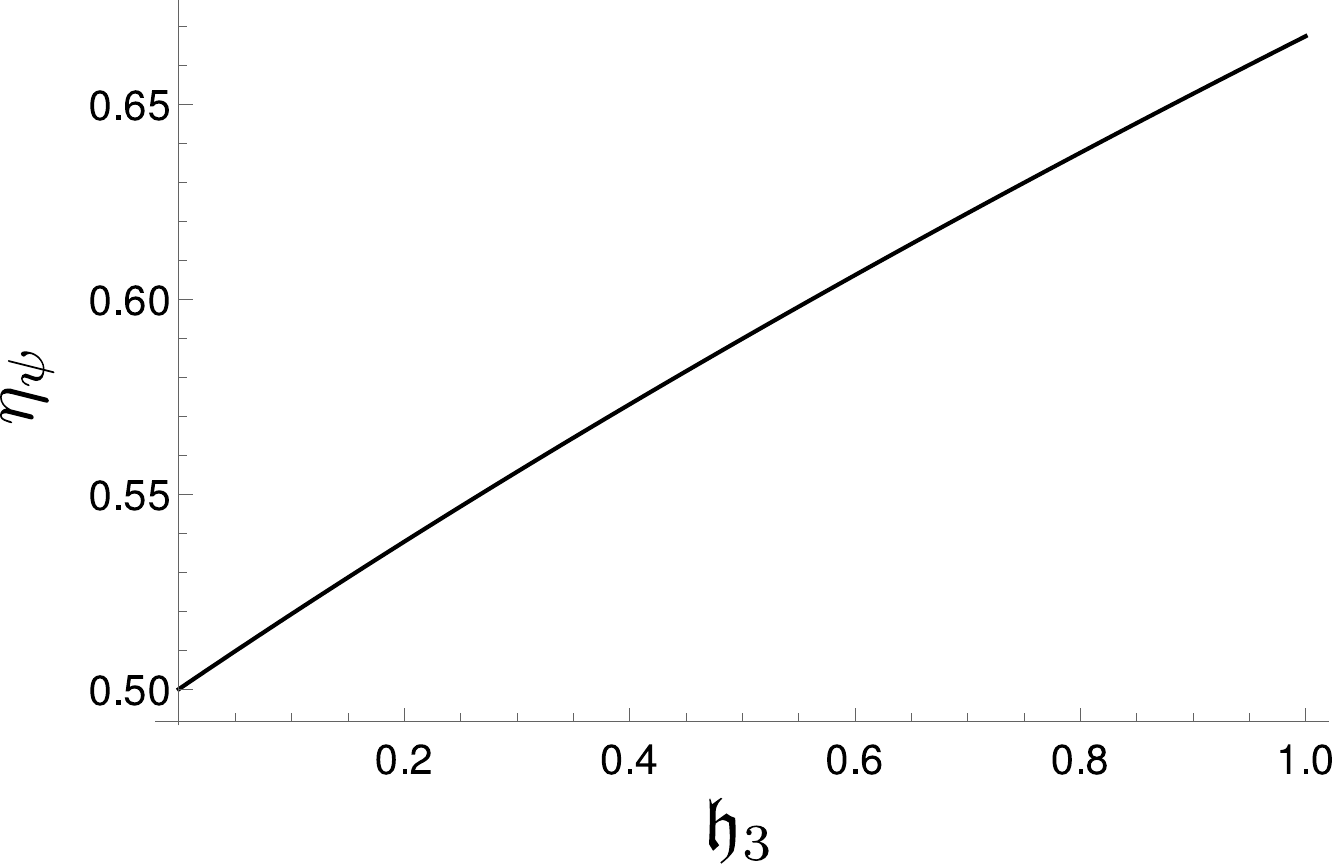}
\caption{The dependence of $\eta_{\psi}$ on $\mf{h}_3$ for the square lattice geometry in $d=3$.
}
\label{fig:eta-psi}
\end{figure}

The fermion propagator on the $\mbf{n}$-th wire is given by
\begin{align}
\avg{\psi_{\mbf{n}}(0,x) \psi_{\mbf{n}}^{\dag}(0,0) }
& \sim \frac{\rho_0}{|\lam_0  x|^{2\eta_{\psi}(\mf h_d)}},
\end{align}
where $\eta_{\psi}(\mf h_d) = \eta_\vtheta(\mf{h}_d) + \eta_\vphi(\mf{h}_d) $, and we have ignored an overall phase factor arising from the correlation between $\vphi$ and $\vtheta$.
In d=2 we obtain $\eta_{\psi}(\mf h_2) = \half \sqrt{1+\mf{h}_2}$, while it is  numerically computed in $d=3$ and its behavior as a function of $\mf h_3$ is shown in \fig{fig:eta-psi}.
In both cases $\eta_\psi > \half$ for $h_d > 0$.
The faster decay of the fermion-fermion correlation compared to the non-interacting limit ($\mf h_d = 0$), implies that \eq{eq:S-2} describes a Luttinger liquid like metallic state in $d>1$.
Indeed this is an example of a  sliding Luttinger liquid state.
Due to the central role played by  inter-fermion Coulomb repulsion, we refer to it as  \emph{Coulombic sliding Luttinger liquid} (CSLL).
Unlike SLLs arising in systems with short-ranged interactions, the CSLL is controlled by a single parameter, $\mf{h}_d$.

\begin{figure}[!t]
\centering
\includegraphics[width=0.85\columnwidth]{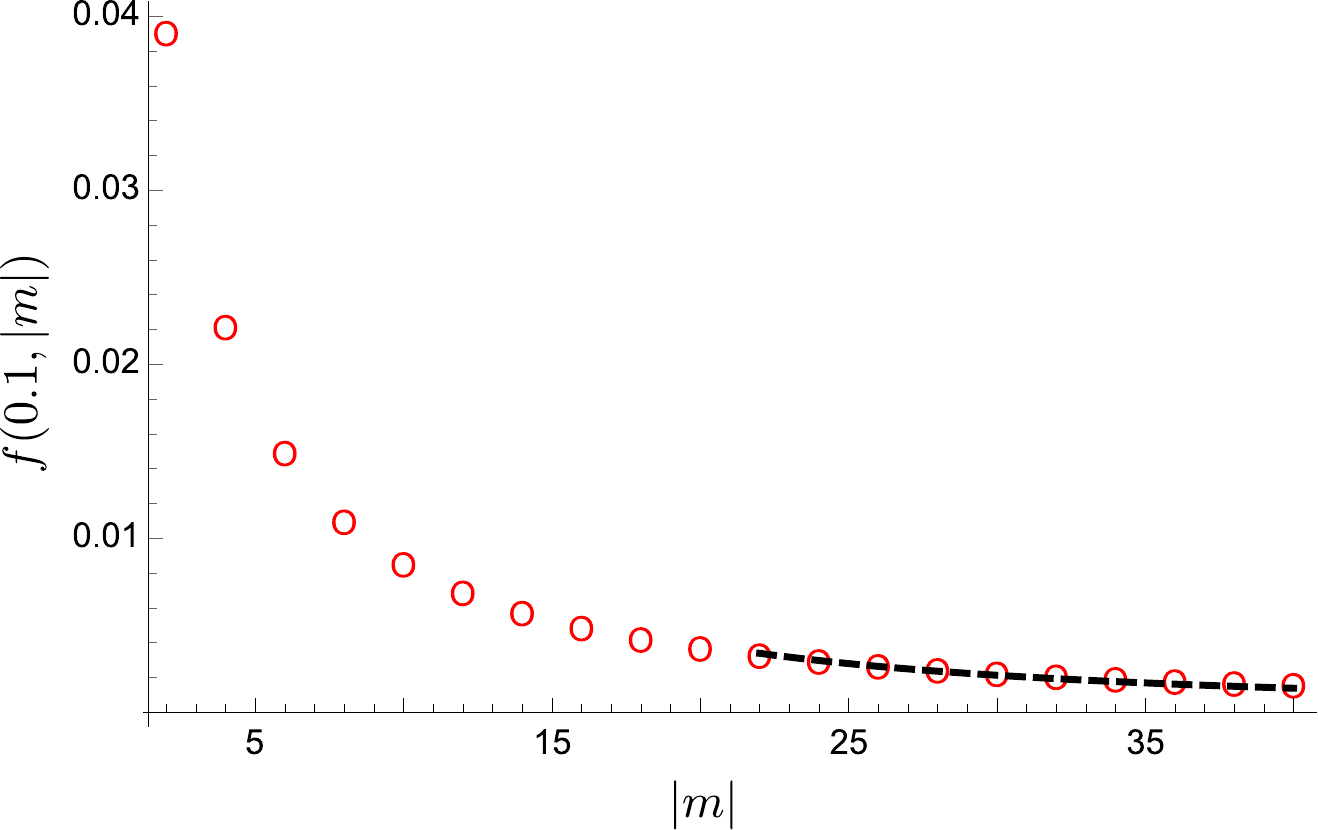}
\caption{Deviation of the interwire density-density correlation at the CSLL fixed point from that at the non-interacting fixed point, as a function of the separation between the wires. Here we have chosen $d=2$ and $\mf h_2 = 0.1$. The empty circles are numerically computed values of $f(0.1,|n-m|)$, and the dashed line is the asymptotic form,  $f(\mf{h}_2,|m|) \sim |m|^{-3/2}$. }
\label{fig:rho-rho}
\end{figure}
Due to the absence of interwire hoppings, the single-particle correlation functions are diagonal in the wire index.
If this were true for all  correlation functions, then the CSLL would be equivalent to a collection of non-interacting LLs, albeit with renormalized exponents.
The distinction is easily demonstrated with the aid of the  density-density correlation between two distinct wires,
\begin{align}
\avg{\rho_{\mbf n}(0, 0) \rho_{\mbf m}(0, 0)} = \rho_0^2 - \lt(\frac{\lam_0}{2\pi}\rt)^2 f(\mf{h}_d,\mbf{n} - \mbf{m}),
\end{align}
where $\mbf{n} \neq \mbf{m}$ and the dimensionless function
\begin{align}
f(\mf{h}_d, \mbf{n} - \mbf{m}) = - \el^{d-1} \int_{BZ} \frac{d^{d-1} K}{(2\pi)^{d-1}} \frac{\cos(\mf{a}  \mbf{K}\cdot(\mbf{n} - \mbf{m}) ) }{\sqrt{1 + \mf{h}_d \lt(\frac{\pi}{\mf{a} | \mbf{K}|} \rt)^{d-1} } }.
\label{eq:f}
\end{align}
We note that $f(\mf{h}_d, \mbf m)$ depends on $\mbf m$ only through $|\mbf m|$.
Due to the cosine factor, the integration over $|\mbf K|$ obtains dominant contribution from the region where $\mf{a} |\mbf K| \lesssim |\mbf n - \mbf m|^{-1}$.
However, the denominator of the integrand suppresses  it at small $|\mbf K|$.
Thus, $f(\mf h_d, |\mbf m|)$    decreases as $|\mbf m|$ increases, and $f(\mf h_d, |\mbf m|) \sim \mf{h}_d^{-1/2} ~ |\mbf m|^{-3(d-1)/2}$ for $|\mbf m| \gg 1$.
This  nontrivial interwire correlation between the densities  is demonstrated in $d=2$ with the aid of  \fig{fig:rho-rho}.
Therefore, the CSLL is distinct from a simple collection of LLs in $d$ dimensions.
While the low energy mode disperses as $\omega = \sqrt{v_F V_\vphi(\mbf K)} |k_x|$ along the wires, the CSLL is a charge insulator in the transverse direction due to a lack of interwire single-particle hopping.

The physics of spinful electrons confined to one dimension, and   interacting through  a three dimensional Coulomb potential was considered by Schulz in   \cite{Schulz93}.
The unscreened tail of the Coulomb interaction was shown to lead to anomalous logarithmic dependences of the correlation functions in Eqs.  \eqref{eq:corr-theta} and \eqref{eq:corr-phi}.
We recover these anomalous logarithms by setting $d=1+\eps$, which leads to
\begin{align}
& \eta_\vtheta(\mf{h}_{1+\eps}) =  \frac{\sqrt{\mf{h}_{1+\eps} + 1} + \mf{h}_{1+\eps} \text{csch}^{-1}\left(\sqrt{\mf{h}_{1+\eps}}\right)}{4\epsilon } \label{eq:eta-theta-1} \\
& \eta_\vphi(\mf{h}_{1+\eps}) =  \frac{\sqrt{\mf{h}_{1+\eps} + 1} -  \mf{h}_{1+\eps} \text{csch}^{-1}\left(\sqrt{\mf{h}_{1+\eps}}\right)}{4\epsilon }.
\label{eq:eta-phi-1}
\end{align}
As $\eps \rtarw 0$ these exponents diverge as $\eps^{-1}$, indicating the presence of additional singularities in $d=1$.
In contrast to the qualitative modification of simple LL behavior by long-range Coulomb interaction in $d = 1$, the properties of the CSLL obtained here are qualitatively similar to those obtained in SLLs with short range interactions.
This difference between $d>1$ and $d=1$ is attributed to interwire-screening, which removes the singularities arising from the long-range tail of the unscreened Coulomb interaction through the enlarged phase space in the transverse direction \cite{Schulz83}.
These conclusions are further supported by diagrammatic computations within the parquet approximation \cite{Barisic83}.
In the earlier works on CSLL in references  \cite{Schulz83,Barisic83,Botric84} the authors focused on the renormalization of the intrawire interactions due to interwire Coulomb interaction, and showed that density wave (pairing) susceptibilities are enhanced (suppressed). Here we have fully characterized both intra- and inter-wire correlations.
In the following section we investigate the effects of interwire backscatterings at the CSLL fixed point.

\section{Instabilities of the Coulombic sliding Luttinger liquid} \label{sec:instabilities}
Weak perturbations that do not break either the sliding symmetry or particle number conservation on each wire will not destabilize the CSLL state.
However, even at low energies there exists processes, viz.   interwire backscatterings and single-particle hoppings, that  break either or both the above  symmetries.
Therefore, it is necessary to consider the effect of these symmetry breaking perturbations on the CSLL fixed point in order to establish the true ground state of the system.
The purpose of this section is to determine the parameter regime, if any, where the CSLL phase is stable, and identify the potential symmetry broken states where it is unstable.

Although the Coulomb interaction primarily  contributes to the forward scattering channel due to the dominance of small-momentum exchange processes, it also mediates weak but non-vanishing $2k_F$ backscatterings.
In general, these backscatterings can destabilize the CSLL phase by utilizing the extensive nesting between the two chiral segments of the Fermi surface in $d > 1$.
The nesting can alter  particle-particle and particle-hole pair hopping amplitudes between different wires.
Further, interwire hoppings are always possible due to non-vanishing single-particle tunneling amplitude between wires.
In order to investigate the stability of the CSLL against these destabilizing tendencies we consider the effect of the following  operators,
\begin{widetext}
\begin{align}
& O_{CDW}^{(\mbf{m})}(\tau, x, \mbf{n})
 = \half \lt[ \psi_{L;\mbf{n}}^{\dag}(\tau, x) \psi_{R;\mbf{n}}(\tau, x) \psi_{R;\mbf{n} + \mbf{m}}^{\dag}(\tau, x)  \psi_{L;\mbf{n} + \mbf{m}}(\tau, x) + \mbox{h.c.} \rt]
=  \rho_0^2 \cos\lt[ 2(\vphi_{\mbf{n}}(\tau, x) - \vphi_{\mbf{n} + \mbf{m}}(\tau, x) )\rt]
\label{eq:cdw} \\
& O_{SC}^{(\mbf{m})}(\tau, x, \mbf{n})
= \half \lt[ \psi_{L;\mbf{n}}^{\dag}(\tau, x) \psi_{R;\mbf{n}}^{\dag}(\tau, x) \psi_{R;\mbf{n} + \mbf{m}}(\tau, x)  \psi_{L;\mbf{n} + \mbf{m}}(\tau, x) + \mbox{h.c.} \rt]
= \rho_0^2 \cos\lt[ 2(\vtheta_{\mbf{n}}(\tau, x) - \vtheta_{\mbf{n} + \mbf{m}}(\tau, x) )\rt],
\label{eq:sc} \\
& O_{sp}^{(\mbf{m})}(\tau, x, \mbf{n})
 = \frac{1}{4} \lt[\psi_{L;\mbf{n}}^{\dag}(\tau, x) \psi_{L;\mbf{n} + \mbf{m}}(\tau, x)
+ \psi_{R;\mbf{n}}^{\dag}(\tau, x) \psi_{R;\mbf{n} + \mbf{m}}(\tau, x) + \mbox{h.c.} \rt] \nn \\
& \qquad \qquad \qquad = \rho_0 \cos\lt[ \vtheta_{\mbf{n}}(\tau, x) - \vtheta_{\mbf{n} + \mbf{m}}(\tau, x) \rt]
\cos\lt[ \vphi_{\mbf{n}}(\tau, x) - \vphi_{\mbf{n} + \mbf{m}}(\tau, x) \rt].
\label{eq:sp}
\end{align}
\end{widetext}
An instability driven by  $\subtxt{O}{CDW}^{(\mbf{m})}$  ($\subtxt{O}{SC}^{(\mbf{m})}$) leads to a CDW (superconducting) state which breaks the continuous sliding symmetry (particle number conservation on each wire) to a discrete symmetry.
The strengthening of $\subtxt{O}{sp}^{(\mbf m)}$ enhances the energy scale for a  crossover from the CSLL state to a $d$ dimensional Fermi liquid metal.
Below the crossover scale both the aforementioned symmetries are broken.
We note that the spinlessness of the fermions does not allow intrawire backscatterings, as a result of which  $\mbf{m} \neq 0$.
Thus, for a fixed wire-stacking,  geometry we obtain a set of operators parameterized by the label $\mbf{m}$, which compete with each other.

The leading instability is identified by comparing the scaling dimensions of  susceptibilities of various operators.
To compute the scaling dimension of the susceptibility of operator $O_X^{(\mbf{m})}$, we perturb the action, \eq{eq:S-2}, with the vertex
\begin{align}
S_X^{(\mbf{m})} = g_X^{(\mbf{m})} \sum_{\mbf{n}} \int \dd{\tau} \dd{x} O_{X}^{(\mbf{m})}(\tau, x, \mbf{n}),
\end{align}
and with the help of the equal time correlation on the $\mbf n$-th wire, $\avg{O_{X}^{(\mbf{m})}(\tau, x, \mbf{n}) O_{X}^{(\mbf{m})}(\tau, x + \Dl x, \mbf{n})} $, we obtain the scaling exponent of $O_{X}^{(\mbf{m})}$.
The scale invariance of $S_X^{(\mbf{m})}$ fixes the  scaling dimension of the coupling $g_X^{(\mbf{m})}$, $ [ g_X^{(\mbf{m})} ] = 2 - [ O_{X}^{(\mbf{m})}]$.
Here the scaling dimension, $[\mc{Y}]$, of an operator $\mc{Y}$ is defined through the relationship $\mc{Y}(\lam) = \mc{Y}(\lam_0) e^{[Y] \ell}$, where $\ell = \ln{(\lam_0/\lam)}$ is the RG time/distance, and $\lam < \lam_0$ is the  running momentum scale.
The asymptotic behavior of the equal-time correlation function of $O_{X}^{(\mbf{m})}$ on the $\mbf n$-th wire at the CSLL fixed point is computed in Appendix \ref{app:sine-gordon}, and it takes the form,
\begin{align}
& \avg{
O_{X}^{(\mbf{m})}(0, \Dl x, \mbf n)
~ O_{X}^{(\mbf{m})}(0,0,\mbf n)
}
= \frac{2 \rho_0^4}{|{\lam_0} \Dl x|^{2 \eta_{X}(\mf{h}_d, \mbf{m}) } }.
\end{align}
On coarse-graining, $g_X^{(\mbf{m})}$ evolves as
\begin{align}
\dow_\ell \wtil{g}_X^{(\mbf{m})}(\ell) = \lt( 2 - \eta_X(\mf{h}_d, \mbf{m}) \rt) \wtil{g}_X^{(\mbf{m})}(\ell),
\label{eq:beta-g}
\end{align}
where $\wtil{g}_X^{(\mbf{m})}(\ell) \equiv \wtil{g}_X^{(\mbf{m})}(\lam) = g_X^{(\mbf{m})} \lam^{-(2-\eta_X(\mf{h}_d, \mbf{m}))}$
is the corresponding dimensionless coupling.
This implies that
the operator $O_X^{(\mbf{m})}$ is a relevant (irrelevant) perturbation at the CSLL fixed point if  $\eta_X(\mf{h}_d, \mbf{m}) < 2$ ($\eta_X(\mf{h}_d, \mbf{m}) > 2$).

We note that, within a weak-coupling framework, generally, symmetry breaking in two and higher dimensional metals is driven by marginal operators.
Consequently, the sign and magnitude of quantum corrections determine the dominant instability.
In the present case, however, we are able to fully account for interactions along the wires, which leads to the interwire operators picking up non-trivial scaling dimensions at tree-level as demonstrated in \eq{eq:beta-g}.
Thus, approaching from weak-coupling side, the dominant instability is determined by the coupling which reaches a order of $1$ value quickest in terms of the RG time, $\ell$.
This also leads to a non-BCS form of various energy scales associated with symmetry breaking transitions in the CSLL metal.

\subsection{Charge density wave instabilities} \label{sec:cdw}
\begin{figure*}[!t]
\centering
 \begin{subfigure}[b]{0.33\textwidth}
\includegraphics[width=0.95\columnwidth]{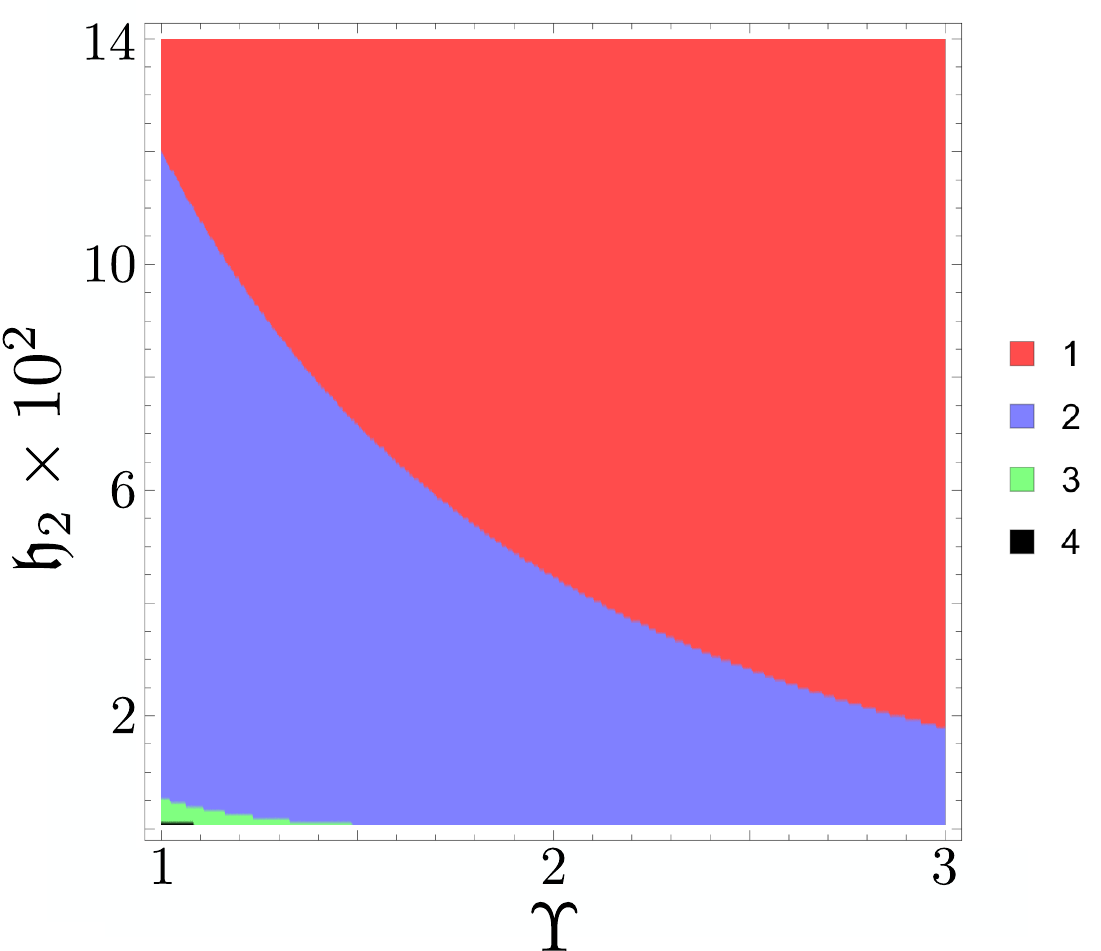}
\caption{}
\label{fig:0T_phase-diag_2d}
 \end{subfigure}
 \begin{subfigure}[b]{0.33\textwidth}
	\includegraphics[width=0.95\columnwidth]{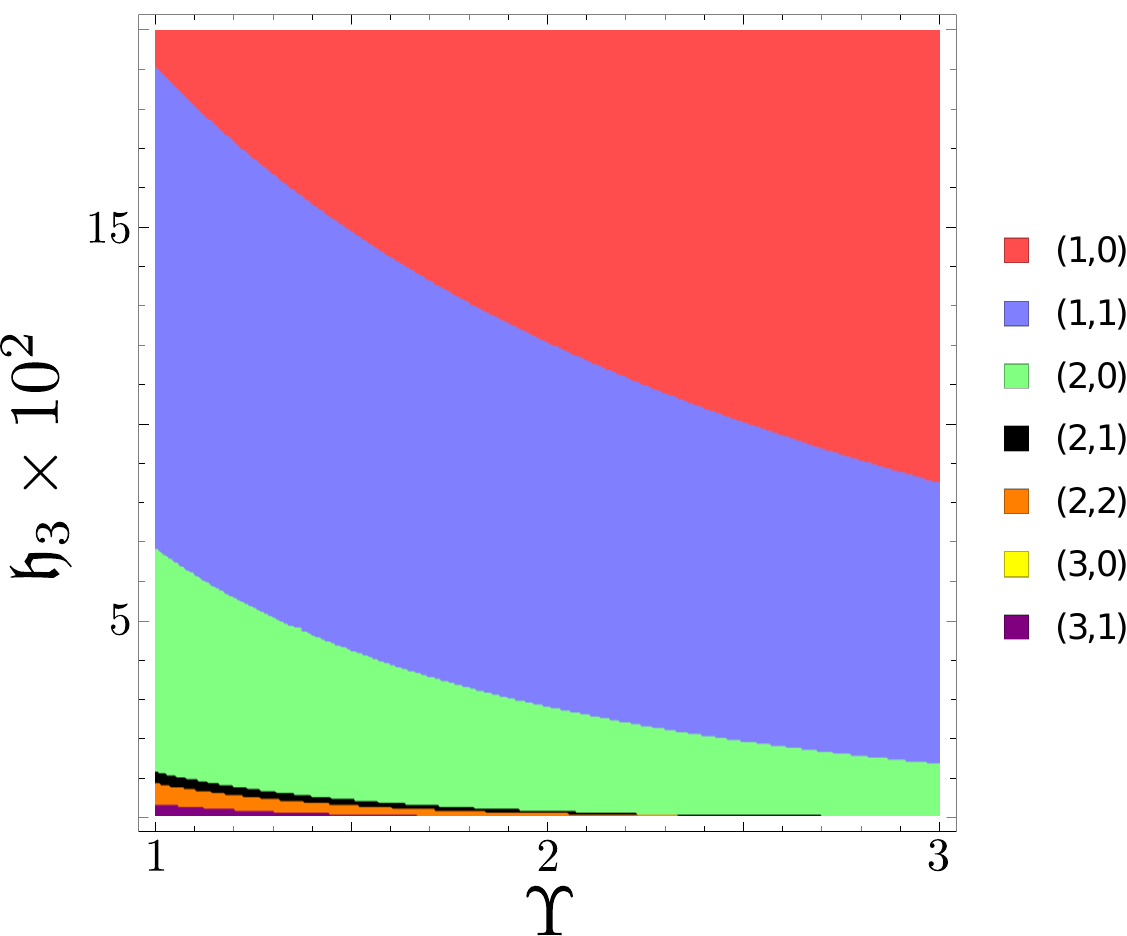}
	\caption{}
	\label{fig:0T_phase-diag_sq}
 \end{subfigure}
 \begin{subfigure}[b]{0.33\textwidth}
	\includegraphics[width=0.95\columnwidth]{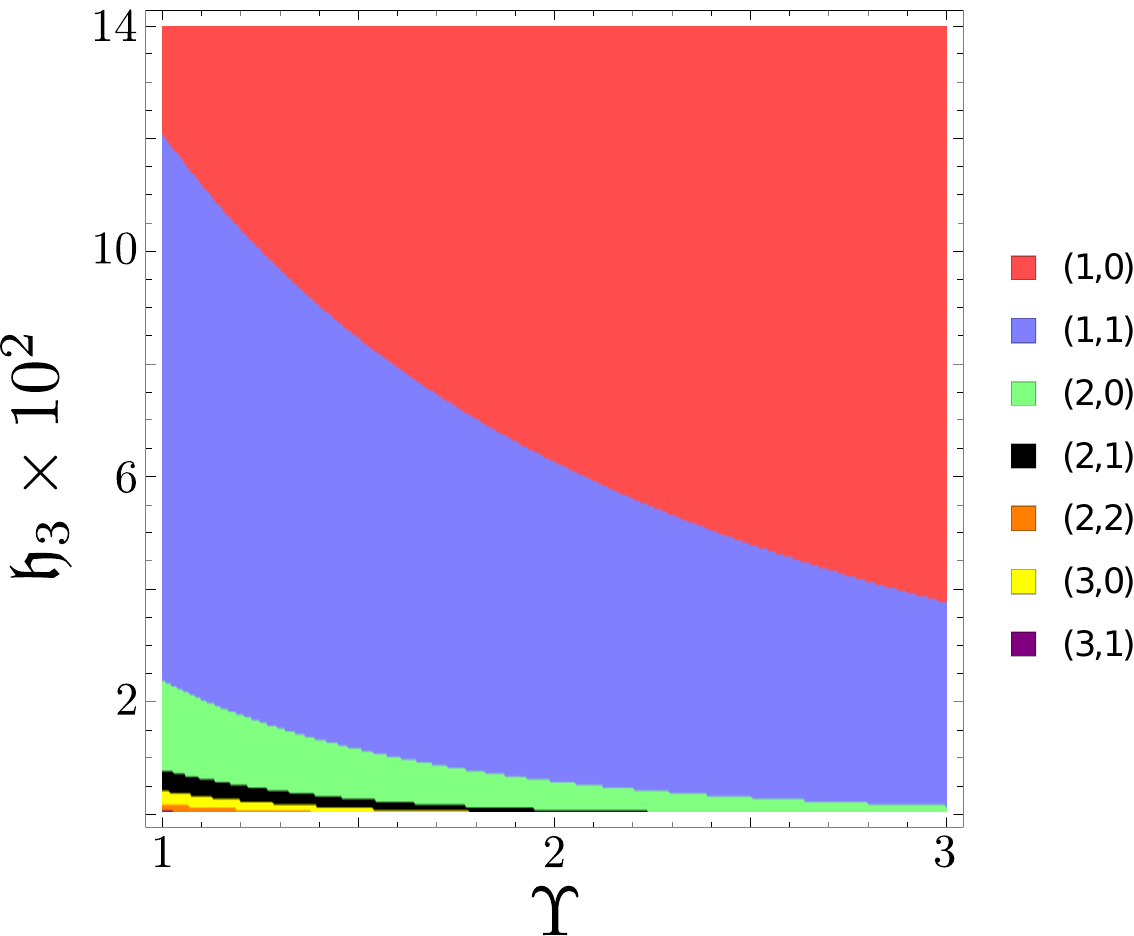}
	\caption{}
	\label{fig:0T_phase-diag_tri}
 \end{subfigure}
\caption{Zero temperature phase diagrams in (a) $d=2$, and $d=3$  with wires stacked in (b) square lattice and (c) triangular lattice geometry.
Here $\mf{h}_d$ is the effective fine structure constant, and  $\Upsilon = 2 k_F \mf{a}$.
The colors represent distinct CDW states which become dominant as $(\mf{h}_d,\Upsilon)$ are tuned.
The legends on the right indicate the direction and periodicity of the ordering vector in the dominant CDW state.
We note that directions related by the $C_n$ point group symmetry of the lattice in $d=3$ are degenerate.
}
\label{fig:phase-diag}
\end{figure*}
As noted in section \ref{sec:SLL}, the repulsive nature of the Coulomb interaction suppresses (enhances) fluctuations in the particle-particle (particle-hole) channel.
This renders the Josephson (SC) couplings irrelevant at the CSLL fixed point, while the CDW couplings become relevant.
In this subsection we consider the effect of the CDW couplings.

As a representative example, let us consider the wires stacked into  a $d-1$ dimensional square lattice.
The scaling dimension of $\subtxt{O}{CDW}^{(\mbf m)}$ is
\begin{align}
& \subtxt{\eta}{CDW}(\mf{h}_d, \mbf{m}) =  2^{2-d} \int_{-1}^{1} d^{d-1} w ~ \frac{ 1 - \cos(\pi \mbf{m} \cdot \mbf{w} ) }{\sqrt{1 + \frac{ \mf{h}_d}{|\mbf{w}|^{d-1} } } }.
\label{eq:eta-cdw-d3-sq}
\end{align}
With the help of \eq{eq:f}, we obtain    $\subtxt{\eta}{CDW}(h_d, \mbf m) = 2\int_0^1 \dd[d-1]{w} (1 + \mf h_d |\mbf w|^{1-d} )^{-1/2} + 2f(\mf h_d, \mbf m)$.
Since the first term is independent of $\mbf m$, $\subtxt{\eta}{CDW}$ follows the trend of $f(\mf h_d, \mbf m)$ as $|\mbf m|$ is tuned.
Thus $\subtxt{\eta}{CDW}$ decreases as $|\mbf m|$ increases.
When $\mf{h}_d > 0$, in the region where $|\mbf w|^{d-1} \ll \mf{h}_d$ the integrand is suppressed by a factor of $\sqrt{|\mbf w|^{(d-1)}}$ compared to the non-interacting fixed point.
Thus, at the CSLL fixed point $\subtxt{\eta}{CDW} < 2$ for any  $\mf h_d, |\mbf m| > 0$, and the fixed point  is \textit{always} unstable to the formation of a CDW in any $d \geq 2$.
We note that, while in $d=2$  $\subtxt{\eta}{CDW}$ is completely determined by $\mf{h}_d$ and $\mbf{m}$,  in  $d > 2$ $\subtxt{\eta}{CDW}$ also depends on the wire-stacking geometry through the cosine term in the numerator of \eq{eq:eta-cdw-d3-sq}.
In Appendix \ref{app:sine-gordon} we demonstrate the stacking geometry dependence by comparing the results for  square and triangular lattices.

In order to identify the leading CDW instability as a function of effective fine structure constant, $\mf{h}_d$, and  lattice geometry, we assume that the coupling $ g_X^{(\mbf{m})}$ results from $2k_F$ backscatterings mediated by Coulomb repulsion, which implies
\begin{align}
\subtxt{\wtil g}{CDW}^{(\mbf{m})}(\mf{h}_d, \Upsilon; \lam_0)
&= \frac{\mf{e}^2}{4 \pi \veps v_F} \int dx \frac{e^{i 2 k_F x}}{\sqrt{x^2 + \mf{a}^2 |\mbf m|^2} } \nn  \\
& \approx \frac{2 \pi^{d-1}}{A_d} ~  \mf{h}_d~ e^{-\Upsilon |\mbf m|},
\label{eq:g0}
\end{align}
where $\Upsilon = 2 k_F \mf{a}$.
We have also assumed that the density of fermions on a given wire is larger than the density of wires, which implies $\Upsilon >  1$.
Equations \eqref{eq:beta-g} and \eqref{eq:eta-cdw-d3-sq} imply that at fixed $\mf{h}_d$ the scaling   dimension of $\subtxt{\wtil g}{CDW}^{(\mbf{m})}$ increases with $|\mbf{m}|$, which suggests that the CDW operator with the largest allowed $|\mbf m|$ drives the dominant instability.
However, the CDW gap, whose magnitude  determines the depth of the free energy minimum, depends on both the scaling dimension and the strength of interwire  backscattering.
In particular, the gap is proportional to  $\subtxt{\wtil g}{CDW}^{(\mbf{m})}(\mf{h}_d, \Upsilon; \lam_0)$, which appears to favor the CDW with $|\mbf m| = 1$.
These opposing tendencies generically lead to a CDW state with a wavevector whose magnitude  lies in between the largest and smallest allowed transverse momenta.
The leading CDW instability at fixed $\mf{h}_d$, $\Upsilon$, and wire-stacking geometry is the one that  \textit{minimizes} the ratio
\begin{align}
\frac{\lam_0}{\subtxt{\lam}{CDW}^{(\mbf m)}(\mf{h}_d, \Upsilon )} =  \lt(\frac{1}{\subtxt{\wtil g}{CDW}^{(\mbf{m})}(\mf{h}_d, \Upsilon; \lam_0)} \rt)^{1/(2 - \eta_{CDW}(\mf{h}_d,\mbf{m}))},
\end{align}
where $\subtxt{\lam}{CDW}^{(\mbf m)}$ is such that $\subtxt{\wtil g}{CDW}^{(\mbf{m})}(\mf{h}_d, \Upsilon; \subtxt{\lam}{CDW}^{(\mbf m)}) \sim 1$.
The algebraic dependence of $\subtxt{\lam}{CDW}^{(\mbf m)}$ on the interwire coupling, 
$\subtxt{\wtil g}{CDW}^{(\mbf{m})}$, is a non-perturbative effect that results from the inclusion of \emph{all} intrawire interactions.
This relationship, however, is subject to the bare  couplings  $ \subtxt{\wtil g}{CDW}^{(\mbf{m})}(\mf{h}_d, \Upsilon; \subtxt{\lam}{0})$ being  small.
The resultant $T=0$ phase diagrams in $d=2 $ and $3$ are shown in \fig{fig:phase-diag}.
We note that for lattices with $C_n$ point group symmetry, CDW states with wavevectors related by the $C_n$ symmetry are degenerate.

From the phase diagrams we deduce that CDW states with larger  wavevectors are favored at stronger interaction strengths and larger interwire spacings.
MacDonald and Fisher identified the CDW state formed between adjacent wires in $d=2$ as a Wigner crystal   \cite{MacDonald00}. 
However, as shown in \fig{fig:phase-diag}, the possible symmetry broken states  extend beyond Wigner crystals, as CDW states with smaller wavevectors and multiple sites per unit cell (in the transverse lattice) are stabilized through the competition between bare coupling strengths and  scaling dimensions of interwire CDW operators.
While the leading instability fixes the ordering wavevector and, consequently, the number of sites per unit cell, it cannot determine the intra-unit cell ordering pattern which is fixed by sub-leading operators. 
We note that the staggered pattern of CDW modulation is most readily realized in CDW states with 2 sites per unit cell.

\begin{figure*}[!t]
\centering
 \begin{subfigure}[b]{\columnwidth}
	\includegraphics[width=0.8\columnwidth]{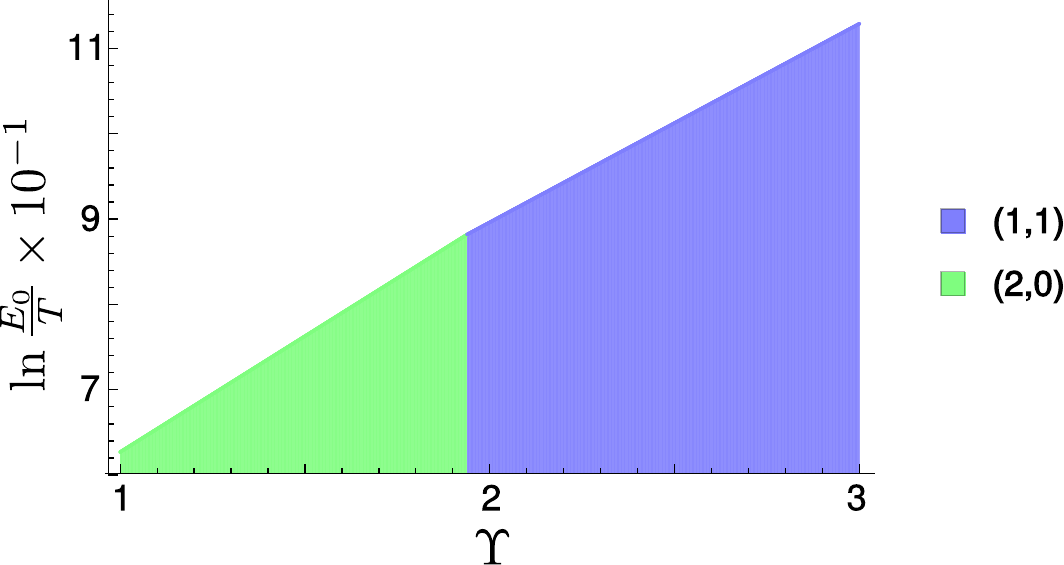}
	\caption{}
	\label{fig:finiteT_phase-diag_h1}
 \end{subfigure}
  \begin{subfigure}[b]{\columnwidth}
 	\includegraphics[width=0.8\columnwidth]{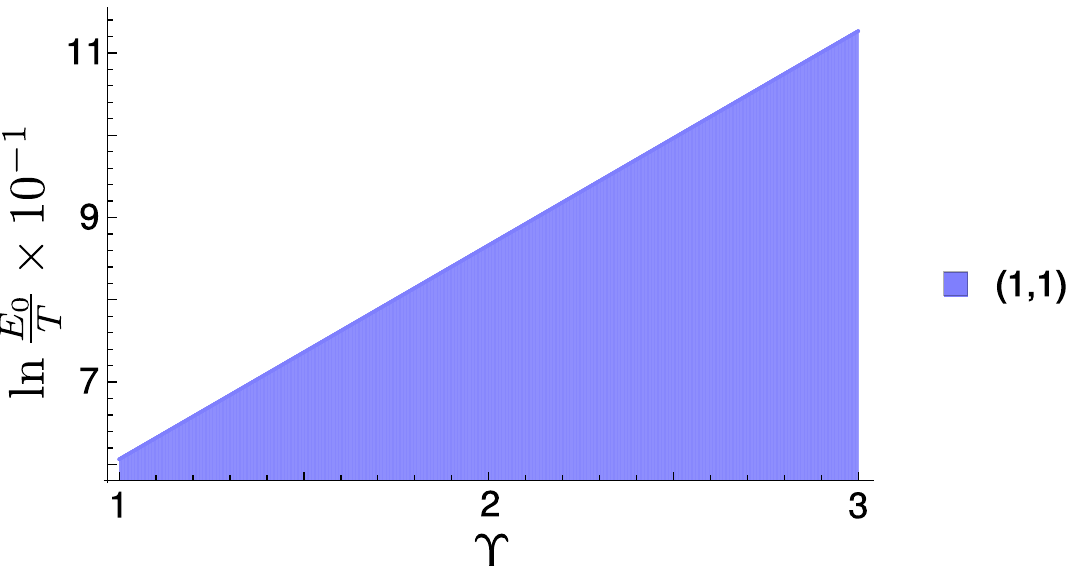}
 	\caption{}
 	\label{fig:finiteT_phase-diag_h2}
  \end{subfigure}
\caption{Finite temperature phase diagram in $d=3$ showing the transition temperature $T_c$ (top boundary of the colored regions) as a function of $\Upsilon = 2 k_F \mf{a}$ for the (a) square, and (b) triangular lattice geometries, respectively.
The colors represent distinct CDW states which become dominant as $\Upsilon$ is tuned.
The legends on the right of each diagram indicate the direction and periodicity of the ordering vector in the dominant CDW state.
Here $E_0 = v_F \lam_0$, and  $\mf{h}_3 = \frac{A_3}{137} \frac{(c/v_F)}{\pi^2 \veps_r}$ with $c / v_F = 100$ and   relative permittivity $\veps_r = 10$.
We note that directions related by the $C_n$ point group symmetry of the lattice are degenerate.}
\label{fig:finiteT_phase-diag}
\end{figure*}
Although the CSLL state is unstable at $T=0$, it exists above a critical temperature,
\begin{align}
T_c^{(\mbf m)}(\mf{h}_d, \Upsilon) \sim v_F \lam_0  \lt(\subtxt{\wtil g}{CDW}^{(\mbf{m})}(\mf{h}_d, \Upsilon; \lam_0) \rt)^{\frac{1}{2 - \eta_{CDW}(\mf{h}_d,\mbf{m})} }.
\label{eq:Tc-SLL}
\end{align}
The corresponding finite-$T$ phase diagrams in $d=3$ are shown in  Figs. \ref{fig:finiteT_phase-diag_h1} and \ref{fig:finiteT_phase-diag_h2} for the square and triangular lattice geometries, respectively.
We note that at the non-interacting fixed point described by \eq{eq:S0-2}   $\subtxt{g}{CDW}^{(\mbf m)}$ is marginal, which implies
\begin{align}
T_c^{(\mbf m)}(\mf{h}_d, \Upsilon) \sim v_F \lam_0 ~ \exp{-\frac{1}{\pha \subtxt{\wtil{g}}{CDW}^{(\mbf m)}(\mf{h}_d, \Upsilon; \lam_0) } },
\label{eq:Tc-bcs}
\end{align}
has the BCS form with $\pha$ being a non-universal numerical factor.
Here the $T_c^{(\mbf m)}$ is solely determined by the strength of interwire Coulomb repulsion.
Since the interaction between nearest-neighbor wires is  strongest, \eq{eq:Tc-bcs} implies that CDW states that modulate over a lattice spacing  is the dominant instability for any $\mf{h}_d$ and $\Upsilon$.
This is in sharp contrast to the result obtained in \eq{eq:Tc-SLL}  by perturbing at the CSLL fixed point with the same operator, where more general CDW states are possible.

\subsection{Dimensional crossover} \label{sec:sp}
While discussing the CDW instabilities, we implicitly assumed that the energy scale below which interwire single particle hoppings become important is small compared to $\subtxt{\lam}{CDW}^{(\mbf m)}$.
The effects of interwire hoppings in quasi-1 dimensional metals have undergone extensive investigations   \cite{Brazovskii85,Bourbonnais91,Castellani92,Boies95,Kopietz97}. 
Here we will estimate the crossover scale, that is accessible within our approach, below which the interwire hoppings cannot be ignored, and the coupled-LL framework becomes inconvenient for describing the physics \cite{Wen90}.
We will show, in particular, that at small $\mf{h}_d$ the interwire hopping amplitudes, $\subtxt{g}{sp}^{(\mbf{m})}$, obtain a larger scaling dimension than the CDW couplings, which implies that the dimensional crossover potentially preempts the CDW instabilities.
However, if the bare  $\subtxt{g}{sp}^{(\mbf{m})} \ll \subtxt{g}{CDW}^{(\mbf{m})}$, the  dimensional crossover scale is pushed below the CDW gap.

The correlation functions of  the interwire hopping operators decay as
\begin{align}
\avg{\subtxt{O}{sp}^{(\mbf m)}(0,  \Dl x, \mbf n) \subtxt{O}{sp}^{(\mbf m)}(0, 0, \mbf n)  } \sim \rho_0 |\lam_0 \Dl x|^{-2\eta_{sp}(\mf{h}_d, \mbf m)},
\end{align}
where $\subtxt{ \eta}{sp}(\mf{h}_d, \mbf m) = (\subtxt{ \eta}{CDW}(\mf{h}_d, \mbf m) + \subtxt{\eta}{SC}(\mf{h}_d, \mbf m))/4$.
By defining $\subtxt{\eta}{CDW}(\mf{h}_d, \mbf m) = 2 -  4\eps_1(\mf h_d, \mbf m)$ and  $\subtxt{\eta}{SC}(\mf{h}_d, \mbf m) = 2 + 4\eps_2(\mf h_d, \mbf m)$,
we express  $\subtxt{\eta}{sp}(\mf{h}_d, \mbf m) = 1 - \eps_1(\mf h_d, \mbf m) + \eps_2(\mf h_d, \mbf m)$, where $\eps_i(\mf h_d, \mbf m) \geq 0$.
Since $\lim_{\mf h_d \rtarw 0} \eps_i(\mf h_d, \mbf m) = 0^+$, weak Coulomb repulsion is not sufficient for overcoming the large bare scaling dimension of $\subtxt{g}{sp}^{(\mbf m)}$.
Consequently, the system may undergo a  dimensional crossover below a momentum scale
\begin{align}
\subtxt{\lam}{cross}^{(\mbf m)} (\Upsilon, \mbf h_d) = \lam_0 \lt( \subtxt{ \wtil g}{sp}^{(\mbf m)} (\Upsilon, \mbf h_d; \lam_0) \rt)^{\frac{1}{1 + \eps_1(\mf h_d, \mbf m) - \eps_2(\mf h_d, \mbf m)} }.
\end{align}
Such a crossover from a lower to a higher dimensional metallic state may be modified by quantum fluctuations  that were not considered in this work. 
Moreover, the higher dimensional metal itself may become unstable to the formation of a density wave or superconducting state \cite{Nickel05}, in which case the crossover would get masked by the symmetry broken state.

Within the purview of the present analysis the CDW transitions discussed in section \ref{sec:cdw} are present if $\subtxt{\lam}{cross}^{(\mbf m)} < \subtxt{\lam}{CDW}^{(\mbf m)}$.
In contrast, for  $\subtxt{\lam}{cross}^{(\mbf m)} > \subtxt{\lam}{CDW}^{(\mbf m)}$ the system crosses over  from the CSLL to a $d$ dimensional Fermi liquid metal which preempts the CDW instabilities.
At fixed $\mbf m$, in the small $\mf h_d$ regime, the former limit is satisfied for $\subtxt{ \wtil g}{sp}^{(\mbf m)} <  \subtxt{ \wtil g}{CDW}^{(\mbf m)} < 1 $.
We note that such a limit is physical, since the interwire hopping and backscattering originate from distinct processes, viz. interwire single-particle tunneling and Coulomb interaction, respectively, as a result of which they are independently tunable.
In particular, the rate of exponential decay of the interwire hopping amplitude  with increasing interwire distance is controlled by a short-distance scale on the order of interwire lattice spacing, while the decay rate of the strength of  backscatterings is controlled by the average interparticle distance on each wire as shown in \eq{eq:g0}.
Since in a dilute system the latter is much smaller than the former, interwire backscatterings dominate over interwire hoppings.

\section{Conclusion} \label{sec:conclusion}
In this paper we analyzed the effect of backscatterings at the CSLL fixed point obtained by coupling an infinite number of quantum wires in $d$ dimensions with Coulomb interaction in the forward scattering channel.
We showed that in the absence of a  dimensional crossover, an infinitesimal Coulomb interaction destabilizes the CSLL towards CDW ordering.
This implies that the metallic state discussed in reference \cite{Zhang16} is likely unstable.
Several CDW states, including Wigner crystals, are shown to compete at linear order in the backscattering couplings.
While CDW states with larger  wavevectors are favored at large values of the effective fine structure constant, $\mf{h}_d$, and low density, CDW states with smaller wavevectors become dominant in the opposite limit.
These properties are demonstrated by constructing both zero and finite temperature phase diagrams.

Upon the inclusion of quantum fluctuations in terms of interwire backscatterings the relevant (in RG sense)  backscattering couplings at quadratic order are expected to renormalize  the transition scales  \cite{Sierra03}.
Furthermore, it is in principle possible to obtain critical fixed points where a subset of the  $\subtxt{g}{CDW}^{(\mbf m)} \propto \mf h_d$.
Both outcomes will modify the phase diagrams obtained here.
Since  there is a large number of relevant couplings with finely spaced scaling dimensions at small $\mf h_d$, a general analysis of the higher order effects is  complicated.
However, by focusing on specific regions of the phase diagram, eg.  fixing $\mf h_d$, the physics may become more amenable to loop-wise renormalization group analyses.
We leave such considerations to future work.


\acknowledgements{
We thank Xi Dai and Victor Quito for helpful discussions. This work was supported by the National Science Foundation (Grants No. DMR-1004545 and No. DMR-1442366).
}

\newpage
\appendix
\setcounter{equation}{0}
\renewcommand{\theequation}{\thesection.\arabic{equation}}
\onecolumngrid
%
\section{Derivation of anomalous dimension of hydrodynamic fields} \label{app:eta}
In this appendix we outline the derivation of the scaling dimensions of the hydrodynamic fields $\vtheta_{\mbf{n}}$ and $\vphi_{\mbf{n}}$.
We consider two different geometries of the first $BZ$, viz. square and triangular lattices.
\subsection{Square lattice}
\label{app:sqBZ}
For a $d-1$ dimensional square lattice, the $BZ$ is a square of sides $\frac{2\pi}{\mf{a}}$.
The correlation functions in Eqs. \eqref{eq:corr-theta} and  \eqref{eq:corr-phi} are given by 
\begin{align}
 \avg{ e^{i \vphi_{\mbf{n}}(\tau, x)}
e^{- i \vphi_{\mbf{n}}(\tau, x + \Dl x)} }
&= \exp{- \mf{a}^{d-1}
\int_{BZ} \frac{\dd[d-1]{K}}{(2\pi)^{d-1}} \int_{-\infty}^{\infty} \frac{dk_0 dk_x}{(2\pi)^{2}}  \lt[1 - \cos(x k_x) \rt] \frac{\pi v_F \Xi_{\lam_0}(k_x)}{k_0^2 + v_F V_\vphi( \mbf{K}) k_x^2}  } 
 = |{\lam_0} x|^{-2\eta_\vphi}\\
\avg{ e^{i  \vtheta_{\mbf{n}}(\tau, x)}
e^{- i \vtheta_{\mbf{n}}(\tau, x + \Dl x)} }
&= \exp{- \mf{a}^{d-1}
\int_{BZ} \frac{\dd[d-1]{K}}{(2\pi)^{d-1}} \int_{-\infty}^{\infty} \frac{dk_0 dk_x}{(2\pi)^2}  \lt[1 - \cos(x k_x) \rt] \frac{\pi V_\vphi( \mbf{K}) \Xi_{\lam_0}(k_x)}{k_0^2 + v_F V_\vphi( \mbf{K}) k_x^2}  } 
 = |{\lam_0} x|^{-2\eta_\vtheta},
\end{align}
where $\Xi_{\lam_0}(k_x)$ is a UV regulator for $k_x$. 
It is convenient to choose a soft cutoff, eg. $\Xi_{\lam_0}(k_x) = \exp{-|k_x|/{\lam_0}}$, since a hard cutoff, $\Xi_{\lam_0}(k_x) = \Theta(\lam_0 - |k_x|)$ with $\Theta(x)$ being the Heaviside theta function, leads to unphysical oscillations.
We note that our choice of cutoff breaks the $1+1$ dimensional   Lorentz invariance  of the Gaussian fixed point.
Therefore, technically this choice is not appropriate, since we do not expect the quantum fluctuations to break the Lorentz invariance. 
However, in the results presented here the absence of Lorentz invariance does not affect the scaling exponents; it only modifies the prefactors of the scaling terms.

The scaling exponents above are given by 
\begin{align}
& \eta_\vphi =  \frac{\mf{a}^{d-1}}{4}
\int_{-\pi/\mf{a}}^{\pi/\mf{a}} \frac{\dd[d-1]{K}}{(2\pi)^{d-1}} \lt[1+\frac{\pi^{d-1} \mf{h}_d}{\mf{a}^{d-1} |\mbf K|^{d-1}} \rt]^{-\half}, \\
& \eta_\vtheta = \frac{\mf{a}^{d-1}}{4}
\int_{-\pi/\mf{a}}^{\pi/\mf{a}} \frac{\dd[d-1]{K}}{(2\pi)^{d-1}} \lt[ 1 + \frac{\pi^{d-1} \mf{h}_d}{\mf{a}^{d-1} |\mbf K|^{d-1}} \rt]^{\half}.
\end{align}
For $\mf{h}_d > 0$, $\eta_\vphi < 0$ and $\eta_\vtheta > 0$, which implies that the phase ($\vtheta$) correlation weakens at large separation, while the density ($\vphi$) correlation is enhanced.


\subsection{Triangular lattice}
\label{app:triBZ}
We repeat the same calculation as for the square lattice for a triangular lattice in $d=3$.
For the lattice in coordinate space we choose the primitive vectors as, $\mbf{e}_1 = \mf{a} \hat y$ and $\mbf{e}_2 = \mf{a} (\hat y/2 + \sqrt{3} \hat z/2)$, where $\mf{a}$ is the lattice spacing.
The reciprocal vectors, $(\mbf{e}_1^*, \mbf{e}_2^*)$, are deduced from the condition $\{\mbf{e}_1^* \cdot \mbf{e}_2 = 0, \mbf{e}_1^* \cdot \mbf{e}_1 = 2\pi \}$ and $\{\mbf{e}_2^* \cdot \mbf{e}_2 = 2\pi, \mbf{e}_2^* \cdot \mbf{e}_1 = 0 \}$.
This leads to $\mbf{e}_1^* = \frac{2\pi}{\mf{a}}(\hat y - \frac{1}{\sqrt{3}} \hat z )$ and $\mbf{e}_2^* = \frac{2\pi}{\mf{a}} \frac{2}{\sqrt{3}} \hat{z} $.
The reciprocal vectors define a hexagonal $BZ$, where the area enclosed by $BZ$ is $\frac{2}{\sqrt{3}} \lt( \frac{2\pi}{\mf{a}}\rt)^2$.
In order for this area to equal $4 \pi^2/\el^2$, we need to choose $\el = \sqrt{\frac{\sqrt{3}}{2}} ~ \mf{a}$.
The sides of the hexagonal $BZ$ is $s = \frac{4\pi}{3\mf{a}}$, and integration of a function, $f(k_y, k_z)$, over $BZ$ is given by
\begin{align}
\mc A[f] = \el^2 \lt[
\int_{-s}^{-\frac{s}{2}} \frac{dk_y}{2\pi} \int_{-\sqrt{3}(k_y + s)} ^{\sqrt{3}(k_y + s)} \frac{dk_z}{2\pi}
+ \int_{-\frac{s}{2}}^{\frac{s}{2}} \frac{dk_y}{2\pi} \int_{-\frac{\sqrt{3}}{2} s} ^{\frac{\sqrt{3}}{2}s} \frac{dk_z}{2\pi}
+ \int_{\frac{s}{2}}^{s} \frac{dk_y}{2\pi} \int_{-\sqrt{3}(s - k_y)} ^{\sqrt{3}(s - k_y)} \frac{dk_z}{2\pi}
\rt] f(k_y, k_z).
\label{eq:int-hex-BZ}
\end{align}
By setting $f(k_y, k_z) = 1$ it is easily checked that $\mc A[1] = 1$, which verifies the required normalization for the integration over $BZ$.

\begin{figure}[!t]
\centering
 \begin{subfigure}[b]{0.4\textwidth}
	\includegraphics[width=0.9\columnwidth]{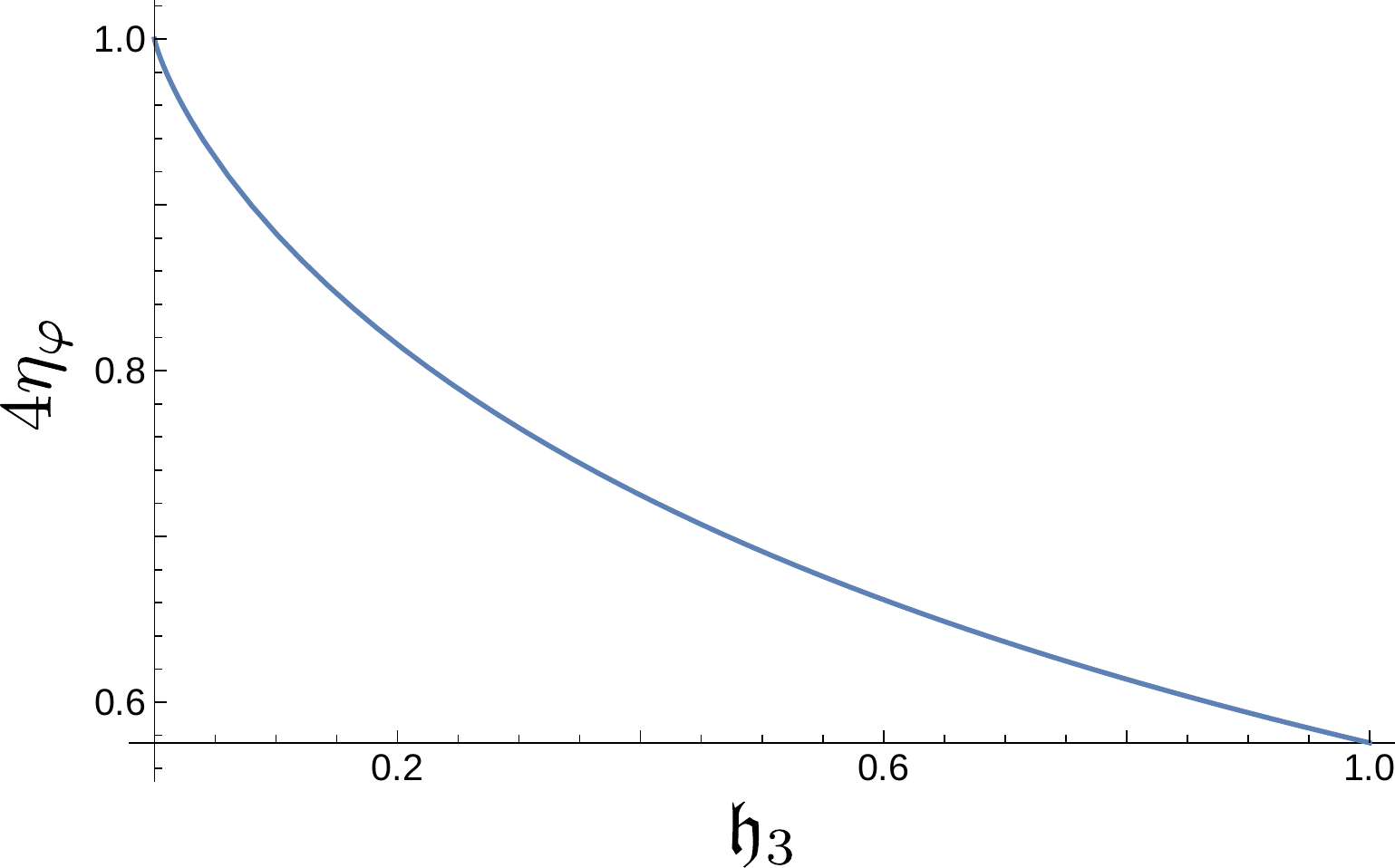}
	\caption{}
	\label{fig:eta-phi-tri}
 \end{subfigure}
 \begin{subfigure}[b]{0.4\textwidth}
	\includegraphics[width=0.9\columnwidth]{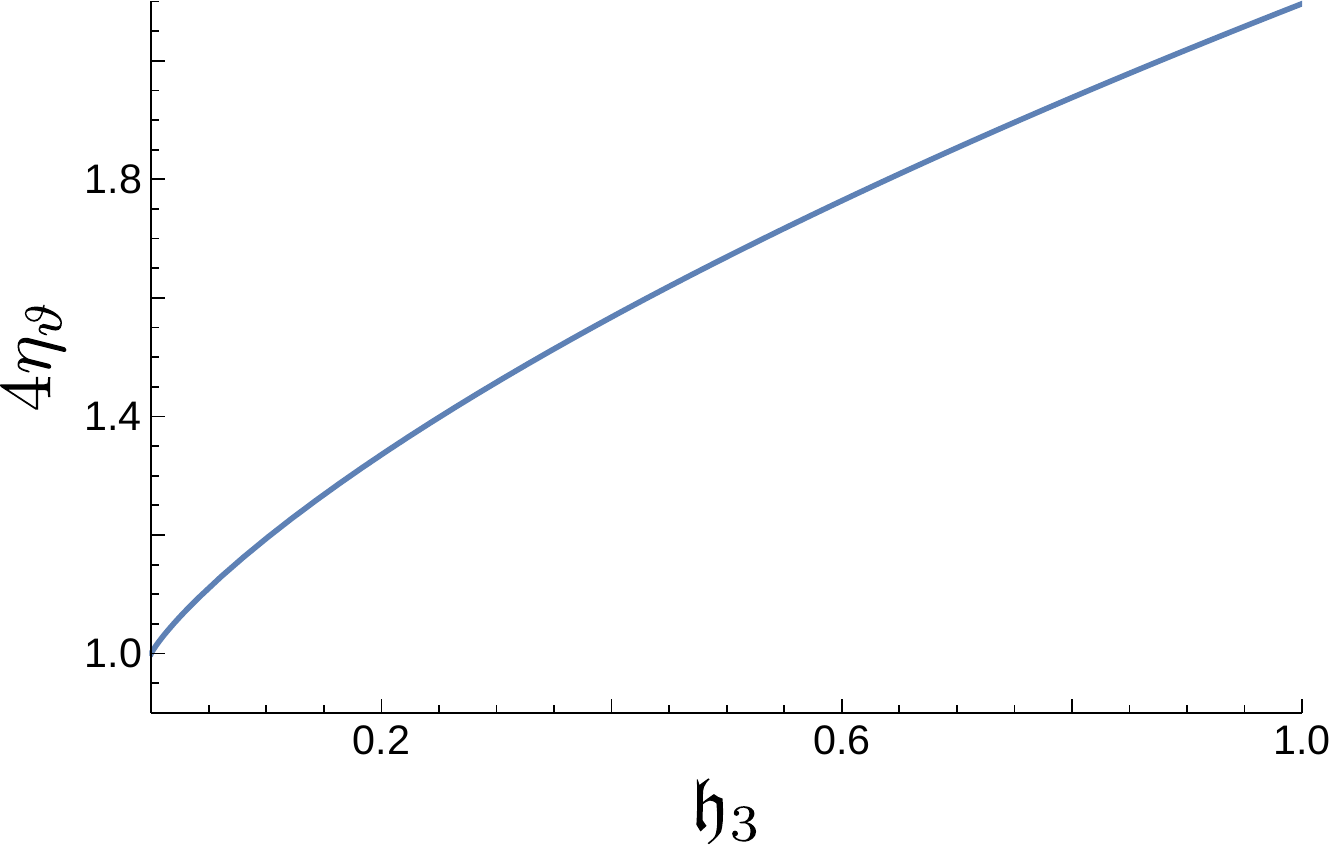}
	\caption{}
	\label{fig:eta-theta-tri}
 \end{subfigure}
\caption{Due to long-range Coulomb interaction the scaling dimension of $e^{i\vphi}$ ($e^{i\vtheta}$) decreases (increases).}
\label{fig:eta-tri}
\end{figure}

The equal-time correlations between the simplest vertex operators are given by,
\begin{align}
& \avg{ e^{i \vtheta_{\mbf{n}}(\tau, x)}
e^{- i \vtheta_{\mbf{n}}(\tau, x + \Dl x)} }
= e^{- \mc{A}\lt[
\int \frac{dk_0 dk_x}{(2\pi)^2} \lt[1 - \cos(x k_x) \rt] \frac{\pi V_\vphi( \mbf{K}) \Xi_{\lam_0}(k_x)}{k_0^2 + v_F V_\vphi( \mbf{K}) k_x^2} \rt] }
= |{\lam_0} x|^{-2\eta_\vtheta}\\
& \avg{ e^{i  \vphi_{\mbf{n}}(\tau, x)}
e^{- i \vphi_{\mbf{n}}(\tau, x + \Dl x)} }
= e^{- \mc{A}\lt[
\int \frac{dk_0 dk_x}{(2\pi)^2} \lt[1 - \cos(x k_x) \rt] \frac{\pi v_F  \Xi_{\lam_0}(k_x)}{k_0^2 + v_F V_\vphi( \mbf{K}) k_x^2} \rt] }
=|{\lam_0} x|^{-2 \eta_\vphi}.
\end{align}
The  exponents are,
\begin{align}
& \eta_\vtheta =
\frac{1}{4} \mc{A}\lt[ \lt( 1+\frac{\pi^2 \mf{h}_3}{\mf{a}^2(k_y^2 + k_z^2)} \rt)^{-1/2} \rt] , \\
& \eta_\vphi = \frac{1}{4} \mc{A}\lt[ \lt( 1+\frac{\pi^2 \mf{h}_3}{\mf{a}^2(k_y^2 + k_z^2)} \rt)^{1/2} \rt] .
\end{align}
Unlike the square $BZ$, it is hard to find analytical expressions for these exponents.
We compute them numerically and plot the results in \fig{fig:eta-tri}.
We check that in the non-interacting limit (i.e. $\mf{h}_3 \rtarw 0$), $\eta_\vtheta = \eta_\vphi = 1/4$.
This reproduces the correct scaling for the left and right moving fermions.



\section{Sine-Gordon terms}
\label{app:sine-gordon}
In this appendix we deduce the scaling dimension of various `sine-Gordon' terms defined in Eqs. \eqref{eq:cdw}, \eqref{eq:sc}, and \eqref{eq:sp}.
We use the methods in Appendix \ref{app:eta} to find 
\begin{align}
& \avg{
O_{CDW}^{(\mbf{m})}(\tau, x, \mbf n)
~ O_{CDW}^{(\mbf{m})}(\tau, x + \Dl x, \mbf n)
}
= \frac{2}{|{\lam_0} \Dl x|^{2 \eta_{CDW}(\mf{h}, \mbf{m}) } } \\
& \avg{
O_{SC}^{(\mbf{m})}(\tau, x,\mbf n)
~ O_{SC}^{(\mbf{m})}(\tau, x + \Dl x,\mbf n)
}
= \frac{2}{|{\lam_0} \Dl x|^{2 \eta_{SC}(\mf{h}, \mbf{m}) } }
\end{align}
The exponents are functions of  $\mf{h}_d$, $\mbf{m}$ and the geometry of the underlying lattice.
For the $d-1$ dimensional square lattice and the triangular lattice they are respectively,
\begin{align}
[\mbox{square}] \qquad & \subtxt{\eta}{CDW}(\mf{h}_d, \mbf{m}) =  \half \int_{-1}^{1} \dd{w_1} \dd{w_2} \lt[ 1 - \cos(\pi \mbf{m} \cdot \mbf{w}) \rt]  \lt(1 + \frac{ \mf{h}_d}{w_1^2 + w_2^2} \rt)^{-\half},
\label{eq:eta-cdw-sq} \\
& \eta_{SC}(\mf{h}_d, \mbf{m}) = \half \int_{-1}^{1} \dd{w_1} \dd{w_2} \lt[ 1 - \cos(\pi \mbf{m} \cdot \mbf{w}) \rt]
 \lt(1 + \frac{ \mf{h}_d}{w_1^2 + w_2^2} \rt)^{\half}
 \label{eq:eta-sc-sq} \\
 [\mbox{triangle}] \qquad  & \subtxt{\eta}{CDW}(\mf{h}_d, \mbf{m}) =  2 \mc{A}\lt[ \lt(1- \cos\lt(\mf{a} \lt( m_1+ \frac{m_2}{2} \rt)k_y + \mf{a} \frac{\sqrt{3}}{2} m_2 k_z \rt) \rt)
\lt( 1+\frac{\pi^2 \mf{h}_3}{\mf{a}^2(k_y^2 + k_z^2)} \rt)^{-1/2} \rt],
\label{eq:eta-cdw-tri} \\
& \eta_{SC}(\mf{h}_d, \mbf{m}) = 2 \mc{A}\lt[ \lt(1- \cos\lt(\mf{a} \lt(m_1+ \frac{m_2}{2} \rt)k_y + \mf{a}\frac{\sqrt{3}}{2} m_2 k_z \rt) \rt)
\lt( 1+\frac{\pi^2 \mf{h}_3}{\mf{a}^2(k_y^2 + k_z^2)} \rt)^{1/2} \rt].
 \label{eq:eta-sc-tri}
\end{align}
Here $\mbf{m} = \sum_{i=1}^{d-1} m_i \mbf{e}_i$, where $\{\mbf{e}_i\}$ are the direct lattice primitive  vectors.
Since the interwire interaction is rotationally symmetric, in $d=3$ the behavior of the scaling dimensions, $\subtxt{\eta}{CDW}$ and $\eta_{SC}$, measured along the line $m_2 = 0$ is  identical to that along any \textit{equivalent} line obtained by rotating $\mbf{e}_1$ by $2\pi/n$ for a lattice with $C_n$ symmetry.
New classes of equivalent lines (directions) are constructed by similar transformations of lines along $(1,1), (1,2), (1,3), \ldots$ directions.



\twocolumngrid


\begin{thebibliography}{99}
\bibitem{note1} An effective description in terms of spinless fermions is appropriate for electrons in the presence of an external magnetic field which leads to spin-polarization.
\end{thebibliography}
\end{document}